\documentclass[proceedings]{JHEP3}
\usepackage{amsfonts}
\usepackage{amsmath}
\usepackage{epsfig,multicol}

\newbox\mybox

\newcommand\fverb{\setbox\mybox=\hbox\bgroup\verb}
\newcommand\fverbdo{\egroup\medskip\noindent\fbox{\unhbox\mybox}\ }
\newcommand\fverbit{\egroup\item[\fbox{\unhbox\mybox}]}
\conference{Integrability of non-Hermitian extensions of CMS-models}
\abstract{We consider non-Hermitian but PT-symmetric extensions of
Calogero models, which have been proposed by Basu-Mallick and Kundu
for two types of Lie algebras. We address the question of whether 
these extensions are meaningful for all remaining Lie algebras (Coxeter groups)
and if in addition one may extend the models beyond the rational case to 
trigonometric, hyperbolic and elliptic models.
We find that all these new models remain integrable, albeit for the
non-rational potentials one requires additional terms in the extension
in order to compensate for the breaking of integrability.}

\title{A note on the integrability of non-Hermitian extensions of
Calogero-Moser-Sutherland models}
\author{Andreas Fring \\
Centre for Mathematical Science, City University\\
Northampton Square, London EC1V 0HB, UK\\
E-mail: \email{A.Fring@city.ac.uk}}

\input{tcilatex}

\begin{document}

\section{Introduction}

Traditionally one considers quantum mechanical models and quantum field
theories associated with Hermitian Hamiltonians, as in general they are
guaranteed to have meaningful energy spectra, lead to conservation of
probability densities under time evolution etc. Despite this apparent need
for the Hamiltonian to be Hermitian, non-Hermitian Hamiltonian systems have
been investigated for some time and found to be physical, as for instance in
the context of level crossing \cite{Rot1,Rot2} and 1+1 dimensional quantum
field theory \cite{Holl,David}. Fairly recent, the observation that the
simple one-particle Hamiltonian with potential term $V=x^{2}(ix)^{\nu }$ for 
$\nu \geq 0$ possesses a real and positive spectrum \cite{Bender:1998ke} has
triggered a sequence of investigations \cite%
{Bender:2002yp,Bender:2002vv,Bender:2003fi,Bender:2003gu,Bender:2003ve,Bender:2003wy,
Bender:2004by,Bender:2004ej,Dorey:2001hi,Mostafazadeh:2001nr,Mostafazadeh:2002id, 
Mostafazadeh:2002hb,Mostafazadeh:2002wg,Mostafazadeh:2002pd,Mostafazadeh:2003iz, 
Mostafazadeh:2003gz,Mostafazadeh:2003qb,Mostafazadeh:2004qh,Mostafazadeh:2004tp, 
Mostafazadeh:2004mx,Znojil:1999qt,Znojil:2000ia,Znojil:2000fr,Levai:2000di,
Znojil:2001ij,Bagchi:2001qu}. One of the outcomes of these studies is the conjecture that Hermiticity is
only a sufficient but not a necessary condition for the spectrum to be real
and positive. Instead, one may simply demand the Hamiltonian to be $\emph{PT}
$-invariant in order to ensure the spectrum to be physical. Inevitably,
non-Hermitian Hamiltonians will give rise to various other kinds of
problems, such as an indefinite metric \cite{Mostafazadeh:2003iz,Znojil}, so
that one is forced to give a proper meaning to unitary evolution etc. This
particular problem is overcome by utilizing a new type of symmetry, which
seems to be always present when the Hamiltonian is $\emph{PT}$-invariant,
and use it to define a new inner-product structure which yields positive
definite norms of the associated quantum states \cite%
{Bender:2002vv,Bender:2004cbr}. Thus these type of non-Hermitian theories
can be made consistent and are not in conflict with concepts of standard
quantum mechanics, but can be regarded as meaningful extensions of them.
Encouraged by these results similar investigations have also been extended
to the realm of quantum field theories \cite%
{Frieder,Bender:2004sa,Bender:2004ss,Bender:2005zz,Bender:2005hf}.

The sole requirement of $\emph{PT}$-invariance allows to include also
various types of momentum dependent terms into the potential of the
Hamiltonian, which previously when demanding Hermiticity would have been
excluded. Such type of models are attractive as they lead to interesting
applications in condensed matter physics, because they are usually of
anyonic nature and exhibit generalized exclusion statistics of Haldane type.
For instance, one has considered extensions of simple harmonic oscillators
of the form $V\sim ixp$ \cite{Swanson,Geyer}, the non-linear Schr\"{o}dinger
equation perturbed by higher spatial dispersions \cite{snir}, double delta
potentials \cite{Pi}, etc. Motivated by this, Basu-Mallick and Kundu \cite%
{Basu-Mallick:2000af} have extended the above mentioned investigations from
one to many-particle systems and proposed a new type of model which
constitutes a non-Hermitian extension of the rational $A_{\ell }$-Calogero
models \cite{Cal2} 
\begin{equation}
\mathcal{H}_{BK}=\frac{p^{2}}{2}+\frac{\omega ^{2}}{2}\sum%
\limits_{i}q_{i}^{2}+\frac{g^{2}}{2}\sum\limits_{i\neq k}\frac{1}{%
(q_{i}-q_{k})^{2}}+i\tilde{g}\sum\limits_{i\neq k}\frac{1}{(q_{i}-q_{k})}%
p_{i}\quad g,\tilde{g}\in \mathbb{R},q,p\in \mathbb{R}^{\ell +1},  \label{BK}
\end{equation}%
where $p_{i}\equiv -i\partial /\partial q_{i}$. Clearly this Hamiltonian is
no longer Hermitian, but its extension remains unchanged when transformed
under a time-reversal and a subsequent parity transformation 
\begin{equation}
\emph{P}:~p_{j}\mapsto -p_{j},~~q_{j}\mapsto -q_{j}\qquad \emph{T}%
:~p_{j}\mapsto -p_{j},~~q_{j}\mapsto q_{j}~~,~~i\mapsto -i,
\end{equation}%
i.e.~the Hamiltonian $\mathcal{H}_{BK}$ is $\emph{PT}$-invariant.
Subsequently, various aspects of the model have been studied \cite%
{Basu-Mallick:2001ce,Basu-Mallick:2003pt,Basu-Mallick:2004ye} and
intriguingly it was found that in these models the exclusion and exchange
parameter differ, unlike in the conventional Calogero models, that is the
case $\tilde{g}=0$, where they are identical.

With regard to the standard Calogero models, there are four conceivable
generalizations for the Hamiltonian $\mathcal{H}_{BK}$. First a fairly
trivial one, a formulation independent of the explicit representation for
the roots of the $A_{\ell }$-Weyl group, second a generalization of the
possible potentials including those which are trigonometric, hyperbolic and
elliptic, third a generalization to Lie algebras (or better Coxeter groups)
other than $A_{\ell }$ and fourth the possibility to include more coupling
constants. A generalization of (\ref{BK}) to Calogero models of $B_{\ell }$%
-type has been studied already in \cite{Basu-Mallick:2001ce}. An \ important
question to answer is whether these extended models remain integrable in a
similar way as their original counterparts or whether the additional term
destroys this valuable property. Despite the fact that the issue of
Hermiticity is mainly relevant in the quantum theory, we investigate here
the classical integrability of these models. Most likely this will also be
important in the quantum theory, as it is well known that in these type of
models the quantum theories inherit often many properties of their classical
counterparts, especially the feature of being integrable or not. The main
purpose of this note is to establish which type of extensions of the
Calogero-Moser-Sutherland (CMS) models \cite{Cal2,Mo,Suth1,Per} preserve
integrability.

\section{Integrability of non-Hermitian PT-invariant extensions of CMS-models%
}

For simplicity we ignore for the time being the confining term in (\ref{BK}%
), that means we set $\omega =0$, and investigate the following
generalization of the Basu-Mallick Kundu model with regard to all four of
the above mentioned possible generalizations 
\begin{equation}
\mathcal{H}=\mathcal{H}_{\text{Cal}}+\mathcal{H}_{\text{\emph{PT}}}=\frac{%
p^{2}}{2}+\frac{1}{2}\sum\limits_{\alpha \in \Delta }g_{\alpha }^{2}V(\alpha
\cdot q)+\frac{i}{2}\sum\limits_{\alpha \in \Delta }\tilde{g}_{\alpha
}f(\alpha \cdot q)(\alpha \cdot p).  \label{HH}
\end{equation}%
Here $\Delta $ is any root system invariant under Coxeter transformations.
We further assume that the potential and the function $f(x)$ in $\mathcal{H}%
_{\text{\emph{PT}}}$ are related as $V(x)=f^{2}(x)$. Besides the rational
case $f(x)=1/x$ considered previously for the $A_{\ell }$ and $B_{\ell }$%
-case \cite{Basu-Mallick:2000af,Basu-Mallick:2001ce}, we also want to
consider the remaining possibilities of the CMS-models, the trigonometric
case $f(x)=1/\sin x$, the hyperbolic case $f(x)=1/\sinh x$ and in particular
the elliptic case $f(x)=1/\func{sn}x$. The Hamiltonian $\mathcal{H}_{\text{%
Cal}}$ in (\ref{HH}) is the usual representation independent, meaning the
roots, formulation of the CMS models. The equality of the last term in (\ref%
{BK}) and the last term in (\ref{HH}) for $\tilde{g}=$ $g_{\alpha }$ is
directly seen when the simple roots of $A_{\ell }$ are expressed in their
standard $(\ell +1)$-dimensional representation, see e.g. \cite{Hum}, $%
\alpha _{i}=\varepsilon _{i}-\varepsilon _{i+1}$ for $1\leq i\leq \ell $,
with $\varepsilon _{i}\cdot \varepsilon _{j}=\delta _{ij}$. Having a
formulation independent of the representation of the roots, we can next
address the question of how many different coupling constants are permitted.
A standard argument is to demand the invariance of the potential under the
action of the Coxeter group. As the Coxeter transformations preserve the
inner product structure, roots of the same length are mapped into each
other, such that the roots can be divided into the two subsets of long and
short roots, $\Delta =\Delta _{s}\cup \Delta _{l}$, which are left invariant
by the Coxeter transformations. This means based on demanding invariance,
the extended models possess also two independent coupling constants like
their CMS counterparts 
\begin{equation}
g_{\alpha }=\left\{ 
\begin{array}{c}
g_{s}\quad \text{for }\alpha \in \Delta _{s} \\ 
g_{l}\quad \text{for }\alpha \in \Delta _{l}%
\end{array}%
\right. \qquad \text{and\qquad }\tilde{g}_{\alpha }=\left\{ 
\begin{array}{c}
\tilde{g}_{s}\quad \text{for }\alpha \in \Delta _{s} \\ 
\tilde{g}_{l}\quad \text{for }\alpha \in \Delta _{l}%
\end{array}%
\right. .
\end{equation}%
Demanding integrability often restricts this choice further, e.g. \cite%
{Per,FK,FM}.

For the rational version of the Calogero models, i.e.~when $f(x)=1/x$, we
note next the crucial property 
\begin{equation}
\eta ^{2}=\alpha _{s}^{2}\tilde{g}_{s}^{2}\sum\limits_{\alpha \in \Delta
_{s}}V(\alpha \cdot q)+\alpha _{l}^{2}\tilde{g}_{l}^{2}\sum\limits_{\alpha
\in \Delta _{l}}V(\alpha \cdot q)\quad \quad \text{with \ }\eta =\frac{1}{2}%
\sum\limits_{\alpha \in \Delta }\tilde{g}_{\alpha }f(\alpha \cdot q)\alpha .
\label{cr}
\end{equation}%
Before using (\ref{cr}), let us first consider an argument to establish that
it actually holds. The identity implies that when computing $\eta ^{2}$ all
terms involving products of the form $f(\alpha \cdot q)f(\beta \cdot
q)(\alpha \cdot \beta )$ for which $\alpha \neq \beta $ cancel each other.
To see this we gather all terms in triplets involving two arbitrary roots $%
\alpha ,\beta $ and a third root which is their sum $\gamma =\alpha +\beta $%
. It may happen though that not all three terms of this type appear in the
product $\eta \cdot \eta $ due to the fact that either $\alpha \cdot \beta
=0 $ or $\beta \cdot \gamma =0$. In that case we can suitable add several of
the missing terms in the hope that overall the additional terms sum up to
zero. It turns out that one may always group the terms conveniently and
cancel them by means of four basic identities. Keeping our discussion
representation independent, these identities can be characterized by the
value of the inner product of the co-roots $\hat{\alpha}=2\alpha /\alpha
^{2} $ and $\hat{\beta}=2\beta /\beta ^{2}$. For $\alpha ,\beta ,\gamma
=\alpha +\beta $ we find the relations 
\begin{eqnarray}
-\frac{\alpha \cdot \alpha }{(\alpha \cdot q)(\beta \cdot q)}+\frac{\alpha
\cdot \gamma }{(\alpha \cdot q)(\gamma \cdot q)}+\frac{\beta \cdot \gamma }{%
(\beta \cdot q)(\gamma \cdot q)} &=&0\quad \quad \quad \text{for }\hat{\alpha%
}\cdot \hat{\beta}=0,  \label{b1} \\
\frac{\alpha \cdot \beta }{(\alpha \cdot q)(\beta \cdot q)}+\frac{\alpha
\cdot \gamma }{(\alpha \cdot q)(\gamma \cdot q)}+\frac{\beta \cdot \gamma }{%
(\beta \cdot q)(\gamma \cdot q)} &=&0\quad \quad \quad \text{for }\hat{\alpha%
}\cdot \hat{\beta}=1,  \label{b2} \\
\frac{\alpha \cdot \beta }{(\alpha \cdot q)(\beta \cdot q)}+\frac{\alpha
\cdot \gamma }{(\alpha \cdot q)(\gamma \cdot q)}+\frac{\beta \cdot \beta }{%
(\beta \cdot q)(\gamma \cdot q)} &=&0\quad \quad \quad \text{for }\hat{\alpha%
}\cdot \hat{\beta}=2,  \label{b3} \\
\frac{\alpha \cdot \beta }{(\alpha \cdot q)(\beta \cdot q)}+\frac{\alpha
\cdot \gamma }{(\alpha \cdot q)(\gamma \cdot q)}-\frac{3\beta \cdot \gamma }{%
(\beta \cdot q)(\gamma \cdot q)} &=&0\quad \quad \quad \text{for }\hat{\alpha%
}\cdot \hat{\beta}=3,  \label{b4}
\end{eqnarray}%
which may be used successively to establish (\ref{cr}). Relation (\ref{b2}),
which applies whenever we have three roots of the same length is most
obvious to use as it involves the terms appearing in the sum when computing
the product $\eta \cdot \eta $ and it is just a matter of grouping the term
together. This involves a non-trivial counting as it requires the precise
knowledge of which inner product of the two roots are non-vanishing and also
the information that after the re-grouping there are no leftovers.
Unfortunately we are not aware of a case independent proof for this.
However, we systematically verified this for many Coxeter groups, and based
on that we assume that (\ref{cr}) holds in general. To sustain this we
present here just some selected examples:

As a representative for root systems involving only roots of one length,
such as all those related to simply laced Lie algebras, we consider the $%
A_{3}$-case. The six positive roots in this case are 
\begin{equation}
\Delta _{A_{3}}^{+}=\{\alpha _{1},\alpha _{2},\alpha _{3},\alpha _{4}=\alpha
_{1}+\alpha _{2},\alpha _{5}=\alpha _{2}+\alpha _{3},\alpha _{6}=\alpha
_{1}+\alpha _{2}+\alpha _{3}\}.
\end{equation}
We abbreviate now $\hat{f}_{i}:=\alpha _{i}/(\alpha _{i}\cdot q)$, $\alpha
^{2}=$ $\alpha _{i}^{2}$ for $1\leq i\leq 6$ and $g^{2}=g_{s}^{2}=g_{l}^{2}$%
. Then using that the only non-vanishing off-diagonal entries in the Cartan
matrix $K_{ij}=2\alpha _{i}\cdot \alpha _{j}/\alpha _{j}^{2}$ are $%
K_{12}=K_{21}=K_{23}=K_{23}=-1$, we compute 
\begin{eqnarray}
\eta ^{2} &=&g^{2}\sum\limits_{k=1}^{6}\frac{\alpha ^{2}}{(\alpha _{k}\cdot
q)^{2}}+\alpha ^{2}g^{2}\left( \hat{f}_{1}\cdot \hat{f}_{2}+\hat{f}_{1}\cdot 
\hat{f}_{4}+\hat{f}_{2}\cdot \hat{f}_{4}+\hat{f}_{2}\cdot \hat{f}_{3}+\hat{f}%
_{2}\cdot \hat{f}_{5}+\hat{f}_{3}\cdot \hat{f}_{5}\right.  \notag \\
&&\ \ \ \left. +\hat{f}_{1}\cdot \hat{f}_{5}+\hat{f}_{5}\cdot \hat{f}_{6}+%
\hat{f}_{1}\cdot \hat{f}_{6}+\hat{f}_{4}\cdot \hat{f}_{3}+\hat{f}_{4}\cdot 
\hat{f}_{6}+\hat{f}_{3}\cdot \hat{f}_{6}\right) \\
&=&g^{2}\sum\limits_{k=1}^{6}\frac{\alpha ^{2}}{(\alpha _{k}\cdot q)^{2}}.
\end{eqnarray}
We organized the last terms already successively into triplets in such a way
that it is easy to see that they all cancel directly by means of (\ref{b2}).

The non-simply laced cases are less straightforward. The simplest example
involving long and short roots with $\hat{\alpha}\cdot \hat{\beta}=2$ is the 
$B_{2}$-case. The four positive roots for this are 
\begin{equation}
\Delta _{B_{2}}^{+}=\Delta _{l}^{+}=\{\alpha _{1},\alpha _{3}=\alpha
_{1}+2\alpha _{2}\}\cup \Delta _{s}^{+}=\{\alpha _{2},\alpha _{4}=\alpha
_{1}+\alpha _{2}\}.
\end{equation}
The $B_{2}$-Cartan matrix has entries $K_{12}=-2$ and $K_{21}=-1$, from
which we compute 
\begin{eqnarray}
\eta ^{2} &=&\sum\limits_{k=1}^{4}\frac{g_{k}^{2}\alpha _{k}^{2}}{(\alpha
_{k}\cdot q)^{2}}+\alpha _{l}^{2}g_{l}g_{s}\left( \hat{f}_{1}\cdot \hat{f}%
_{2}+\hat{f}_{1}\cdot \hat{f}_{4}+\hat{f}_{2}\cdot \hat{f}_{3}+\hat{f}%
_{3}\cdot \hat{f}_{4}\right)  \label{1} \\
&=&\sum\limits_{k=1}^{4}\frac{g_{k}^{2}\alpha _{k}^{2}}{(\alpha _{k}\cdot
q)^{2}}+\alpha _{l}^{2}g_{l}g_{s}\left( -\frac{\alpha _{2}^{2}}{(\alpha
_{2}\cdot q)(\alpha _{4}\cdot q)}+\frac{\alpha _{2}^{2}}{(\alpha _{2}\cdot
q)(\alpha _{4}\cdot q)}\right)  \label{2} \\
&=&\sum\limits_{k=1}^{4}\frac{g_{k}^{2}\alpha _{k}^{2}}{(\alpha _{k}\cdot
q)^{2}}.
\end{eqnarray}
Here we used (\ref{b3}) to obtain the penultimate term in (\ref{2}) from the
third and fourth last term in (\ref{1}) and identity (\ref{b1}) to obtain
the last term in (\ref{2}) from the last two terms in (\ref{1}). Overall the
required terms to complete the identities (\ref{b1}), (\ref{b3}) add up to
zero.

Identity (\ref{b3}) is required for the $G_{2}$-case. The six positive roots
are now 
\begin{eqnarray}
\Delta _{G_{2}}^{+} &=&\Delta _{s}^{+}\cup \Delta _{l}^{+} \\
&=&\{\alpha _{1},\alpha _{3}=\alpha _{1}+\alpha _{2},\alpha _{4}=2\alpha
_{1}+\alpha _{2}\}\cup \{\alpha _{2},\alpha _{5}=3\alpha _{1}+\alpha
_{2},\alpha _{6}=3\alpha _{1}+2\alpha _{2}\}.~~~  \notag
\end{eqnarray}
The $G_{2}$-Cartan matrix has entries $K_{12}=-1$ and $K_{21}=-3$, which
yields 
\begin{eqnarray}
\eta ^{2} &=&\sum\limits_{k=1}^{6}\frac{g_{k}^{2}\alpha _{k}^{2}}{(\alpha
_{k}\cdot q)^{2}}+\alpha _{s}^{2}g_{s}^{2}\left( \hat{f}_{1}\cdot \hat{f}%
_{3}+\hat{f}_{1}\cdot \hat{f}_{4}+\hat{f}_{3}\cdot \hat{f}_{4}\right)
+\alpha _{l}^{2}g_{l}^{2}\left( \hat{f}_{2}\cdot \hat{f}_{5}+\hat{f}%
_{2}\cdot \hat{f}_{6}+\hat{f}_{5}\cdot \hat{f}_{6}\right)  \notag  \label{g1}
\\
&&+\ \alpha _{l}^{2}g_{s}g_{l}\left( \hat{f}_{1}\cdot \hat{f}_{2}+\hat{f}%
_{2}\cdot \hat{f}_{3}+\hat{f}_{1}\cdot \hat{f}_{5}+\hat{f}_{4}\cdot \hat{f}%
_{5}+\hat{f}_{3}\cdot \hat{f}_{6}+\hat{f}_{4}\cdot \hat{f}_{6}\right)
\label{g2} \\
&=&\sum\limits_{k=1}^{6}\frac{g_{k}^{2}\alpha _{k}^{2}}{(\alpha _{k}\cdot
q)^{2}}-3\alpha _{l}^{2}g_{s}g_{l}\left( \hat{f}_{1}\cdot \hat{f}_{3}+\hat{f}%
_{1}\cdot \hat{f}_{4}+\hat{f}_{3}\cdot \hat{f}_{4}\right)  \label{g3} \\
&=&\sum\limits_{k=1}^{6}\frac{g_{k}^{2}\alpha _{k}^{2}}{(\alpha _{k}\cdot
q)^{2}}.
\end{eqnarray}
Here we employed (\ref{b2}) to cancel the last two triplets in (\ref{g1}).
In the step from (\ref{g2}) to (\ref{g3}) we used (\ref{b4}) and then cancel
the last three terms in (\ref{g2}) by means of (\ref{b2}).

In a similar manner as for the presented examples we may establish (\ref{cr}%
) for the remaining cases. Unfortunately, we are not aware of a case
independent argument to prove this in complete generality. Furthermore, we
note that (\ref{cr}) does not hold in general for the non-rational
potentials.

We return now to our main line of argument and employ the identity (\ref{cr}%
) to re-write the Hamiltonian (\ref{HH}) as a conventional Calogero model
with shifted momenta 
\begin{equation}
\mathcal{H}=\frac{1}{2}(p+i\eta )^{2}+\frac{1}{2}\sum\limits_{\alpha \in
\Delta }\hat{g}_{\alpha }^{2}V(\alpha \cdot q).  \label{hhh}
\end{equation}%
together with some re-defined coupling constants 
\begin{equation}
\hat{g}_{\alpha }^{2}=\left\{ 
\begin{array}{c}
g_{s}^{2}+\alpha _{s}^{2}\tilde{g}_{s}^{2}\qquad \text{for }\alpha \in
\Delta _{s} \\ 
g_{l}^{2}+\alpha _{l}^{2}\tilde{g}_{l}^{2}\qquad \text{for }\alpha \in
\Delta _{l}%
\end{array}%
\right. .
\end{equation}%
Next we recall \cite{Lax} that classical integrability may be established by
formulating Lax pair operators $L$ and $M$ as functions of the dynamical
variables $q_{i}$ and $p_{i}$, which satisfy the Lax equation $\dot{L}=\left[
L,M\right] $, upon the validity of the classical equation of motion
resulting from the corresponding Hamiltionian. Taking the observation on
board that the extended model and the ordinary Calogero model only differ by
a specific shift in the momenta and a re-definition of the coupling
constants, it is straightforward to see that this also holds for the Lax
operators. Thus we take the conventional Lax operators for the CMS models
and simply replace $p\rightarrow p+i\eta $. One may then check directly that 
\begin{equation}
L=(p+i\eta )\cdot H+i\sum\limits_{\alpha \in \Delta }\hat{g}_{\alpha
}f(\alpha \cdot q)E_{\alpha }\quad \text{and}\quad M=m\cdot
H+i\sum\limits_{\alpha \in \Delta }\hat{g}_{\alpha }f^{\prime }(\alpha \cdot
q)E_{\alpha }  \label{LM}
\end{equation}%
fulfills the Lax equation with the constraint 
\begin{equation}
\dot{q}_{j}=\frac{\partial \mathcal{H}}{\partial p_{j}}=p_{j}+i\eta
_{j}\qquad \text{and\qquad }\dot{p}_{j}=-\frac{\partial \mathcal{H}}{%
\partial q_{j}}=-\frac{\partial \mathcal{H}_{\text{Cal}}}{\partial q_{j}}-i%
\dot{\eta}_{j},
\end{equation}%
if $\dot{L}_{\text{Cal}}=\left[ L_{\text{Cal}},M_{\text{Cal}}\right] $. We
choose here as convention the Cartan-Weyl basis commutation relations 
\begin{equation}
\left[ H_{i},H_{j}\right] =0,~~~\left[ H_{i},E_{\alpha }\right] =\alpha
^{i}E_{\alpha },~~~\left[ E_{\alpha },E_{-\alpha }\right] =\alpha \cdot H,~~~%
\left[ E_{\alpha },E_{\beta }\right] =\varepsilon _{\alpha ,\beta }E_{\alpha
+\beta }.  \label{comm}
\end{equation}%
which is compatible with $\func{tr}(H_{i}H_{j})=\delta _{ij}$, $\func{tr}%
(E_{\alpha }E_{-\alpha })=1$. The vector $m$ can be specified as a function
of the structure constants $\varepsilon _{\alpha ,\beta }$ and the
potential. As all additional terms resulting from the shift $i\eta $ cancel
in the Lax equation, the requirements imposed by integrability, i.e.~the
validity of the Lax equation are exactly the same as in the non-extended
models. Note that the Lax equation is solved directly only for the $A_{\ell
} $-algebra, but for other algebras we have to follow the reduction
procedures as indicated for instance in \cite{OP6,Per,FK,FM}. Having now
established that $L$ and $M$ in (\ref{LM}) are meaningful Lax operators for
the extended Hamiltonian (\ref{hhh}), we may compute backwards and expand
the kinetic term such that we simply obtain 
\begin{equation}
\mathcal{H}=\frac{1}{2}p^{2}+\frac{1}{2}\sum\limits_{\alpha \in \Delta }\hat{%
g}_{\alpha }^{2}V(\alpha \cdot q)+i\eta \cdot p-\frac{1}{2}\eta ^{2}.
\label{HHH}
\end{equation}%
Note that we did not make any assumption on the potential, such that (\ref%
{HHH}) is a non-Hermitian \emph{integrable} extension for CMS models for all
Coxeter groups, including besides the rational also trigonometric,
hyperbolic and elliptic potentials. We observe that when one wishes to
preserve integrability for the non-rational potentials one can not simply
extend the models by the term $i\eta \cdot p$, but one also has to add the
momentum independent term $-\eta ^{2}/2$ in order to compensate for the
integrability breaking effect of that term.

\section{Conclusions}

We have demonstrated that the non-Hermitian extensions for the rational
Calogero model proposed by Basu-Mallick and Kundu for the $A_{\ell }$ and $%
B_{\ell }$ Coxeter groups can be generalized to all remaining groups in such
a way that they are classically integrable. The identity (\ref{cr}) is
crucial in this context and it would be interesting to have a rigorous
generic, i.e. case independent proof for it. The identity for $\eta ^{2}$
ensures that the extended Hamiltonian differs from the original Calogero
model only by the one term $i\eta \cdot p$. This simplicity can only be
maintained when the potential is rational. However, adding one more term as
proposed in (\ref{HHH}) one obtains integrable extensions for all
CMS-models. One should stress that the above argument does not exclude yet
the possibility that (\ref{HHH}) might be integrable even for non-rational
potentials when the last term is dropped. Nonetheless, it establishes that
when we include this term they are integrable for sure. It would be very
interesting to carry out further studies on these new models along the lines
previously followed in \cite%
{Basu-Mallick:2001ce,Basu-Mallick:2003pt,Basu-Mallick:2004ye} and beyond.

Furthermore, it would be interesting to extend the analysis in \cite%
{Brihaye:2003dc}, where the question of solvability ($\neq $ integrability)
of some of the discussed models has been addressed.

\medskip

\noindent \textbf{Acknowledgments}. Discussions with Tanaya Bhattacharyya
and Hugh Jones are gratefully acknowledged.


\end{document}